\pdfoutput=1

 \documentclass[final,authoryear,3p,times]{elsarticle}
\usepackage{amssymb}
\usepackage[latin1]{inputenc}
\usepackage{hyperref}
\usepackage{amstext}
\usepackage[nearskip,caption=false,font=footnotesize]{subfig}

\usepackage{color}
\definecolor{greencolor}{rgb}{0,0.5,0.2}
\definecolor{redcolor}{rgb}{.7,0.,0.}
\definecolor{bluecolor}{rgb}{0,0.,1.}
\definecolor{greycolor}{rgb}{.5,.5,.5}

\journal{Journal of Informetrics}

\begin{document}

\begin{frontmatter}



\title{On time-varying collaboration networks}


\author{Matheus P. Viana\footnote{Corresponding author: vianamp@gmail.com}}
\address{Developmental and Cell Biology\\
University of California Irvine \\
Irvine, California, United States\\}

\author{Diego R. Amancio}
\address{Institute of Physics of S\~ao Carlos\\
University of S\~ao Paulo, P. O. Box 369, Postal Code 13560-970 \\
S\~ao Carlos, S\~ao Paulo, Brazil \\}

\author{Luciano da F. Costa}
\address{Institute of Physics of S\~ao Carlos\\
University of S\~ao Paulo, P. O. Box 369, Postal Code 13560-970 \\
S\~ao Carlos, S\~ao Paulo, Brazil \\}

\begin{abstract}
The patterns of scientific collaboration have been frequently investigated in terms of complex networks without reference to time evolution.  In the present work, we derive collaborative networks (from the arXiv repository) parameterized along time.  By defining the concept of affine group, we identify several interesting trends in scientific collaboration, including the fact that the average size of the affine groups grows exponentially, while the number of authors increases as a power law. We were therefore able to identify, through extrapolation, the possible date when a single affine group is expected to emerge. Characteristic collaboration patterns were identified for each researcher, and their analysis revealed that larger affine groups tend to be less stable.
\end{abstract}


\end{frontmatter}

% \linenumbers

\section{Introduction}

The progressive and inexorable informatization of scientific publishing has implied several important consequences, including the possibility to quantify and analyze the patterns characterizing scientific collaborations. For instance, many efforts have been dedicated to the identification of citations between articles (e.g.~\citep{model,goodpractices,assess,onthe,person,chen}). Another well-developed approach involves mapping and studying collaborations between researchers (e.g.~\citep{hsiang,shrum,newman2,newman1}). Such works are often done by using complex networks~\citep{ref17}. In the case of collaboration networks, each researcher is mapped as a node, while the joint authorships establish the links between those nodes. However, most such efforts disregards time, in the sense that the citation and collaborations are taken along long periods of time. By doing so, important information about transient patterns of collaboration are overlooked. For instance, some collaborations are more likely to follow an intermittent pattern, while others would be expected to proceed along continuous periods of time.

The current work aims precisely at addressing this important issue, which has been accomplished by parameterizing the collaboration networks explicitly along time.  So, instead of a single network, we derive a sequence of networks defined from a starting time up to the present moment (i.e. our networks are cumulative). For each node $i$ in each of such parameterized networks, we define its respective \emph{affine group}, corresponding to two sets of nodes. First, we identify those nodes that are directly attached to $i$, as they are co-authors. The second set of nodes corresponds to those that belong to the same community~\citep{girvan} as node $i$, and therefore represents those authors that are more closely interrelated. Having obtained the time-parameterized networks and the respective affine groups, we proceed to analyze the evolution of the latter along time.  More specifically, we calculate the mean size of the affine groups along time for three different collaboration networks extracted from the arXiv repository (\url{www.arXiv.org}). Remarkably, we found that these sizes scale as an exponential with different exponents, while the number of authors in the respective networks grows slower, as a power law. We also found that different affine groups tend to exhibit rather distinct intermittence patterns, which suggested a classification of the authors according to their time-dependent collaboration patterns. So, for each author, we calculated the maximum size of the affine groups to which it belonged, as well as the average duration of the respective collaborations. These findings suggest that authors who collaborate with more people also tend to have shorter collaborations.

\section{Methodology}

\subsection{The time-varying collaboration network}

\begin{figure} [!Htb]
    \centering
    \subfloat[]
    {{\includegraphics[angle=0, width=0.33\textwidth]{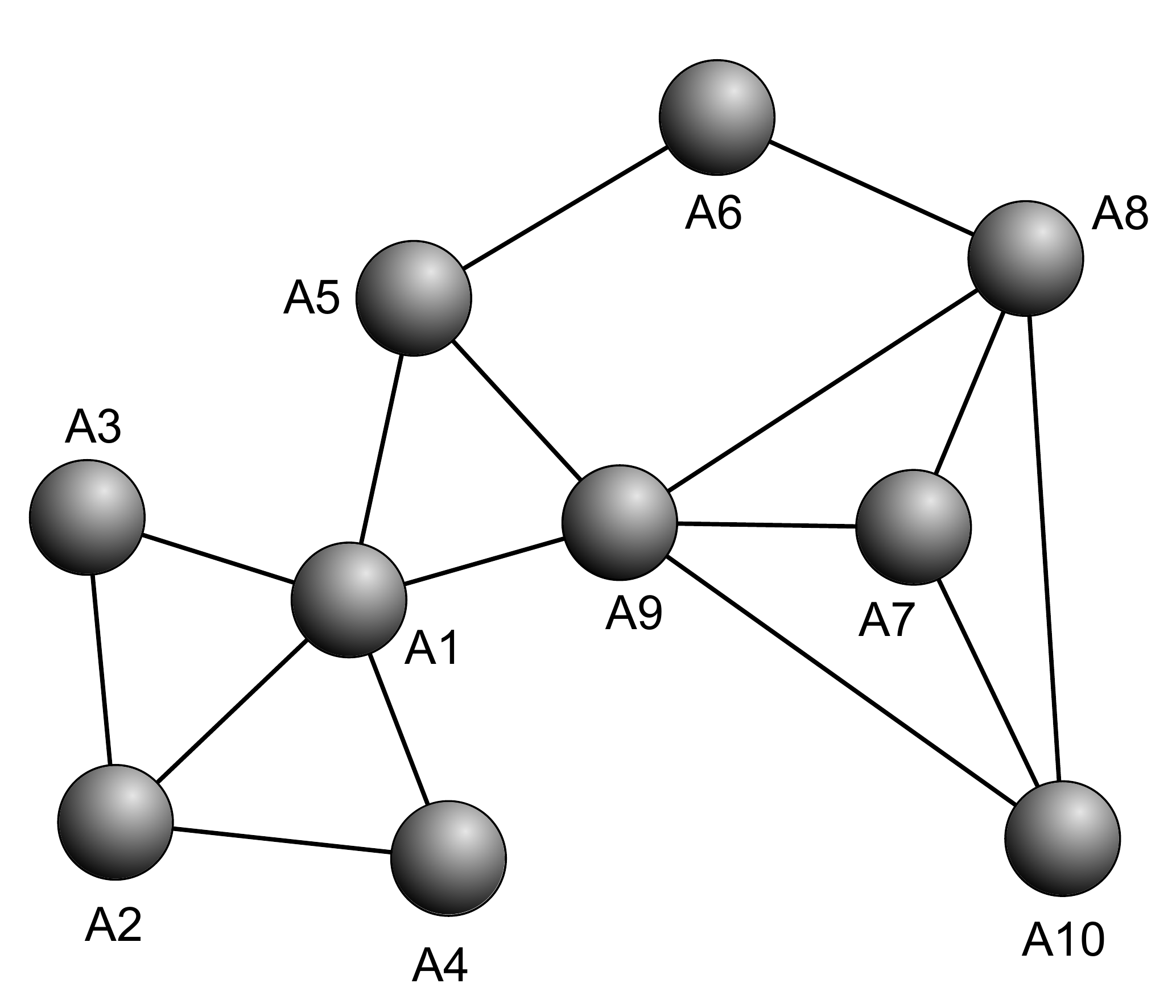}\label{fig:Regular-Lattice-lambda0}}}%
    \subfloat[]
    {{\includegraphics[angle=0, width=0.33\textwidth]{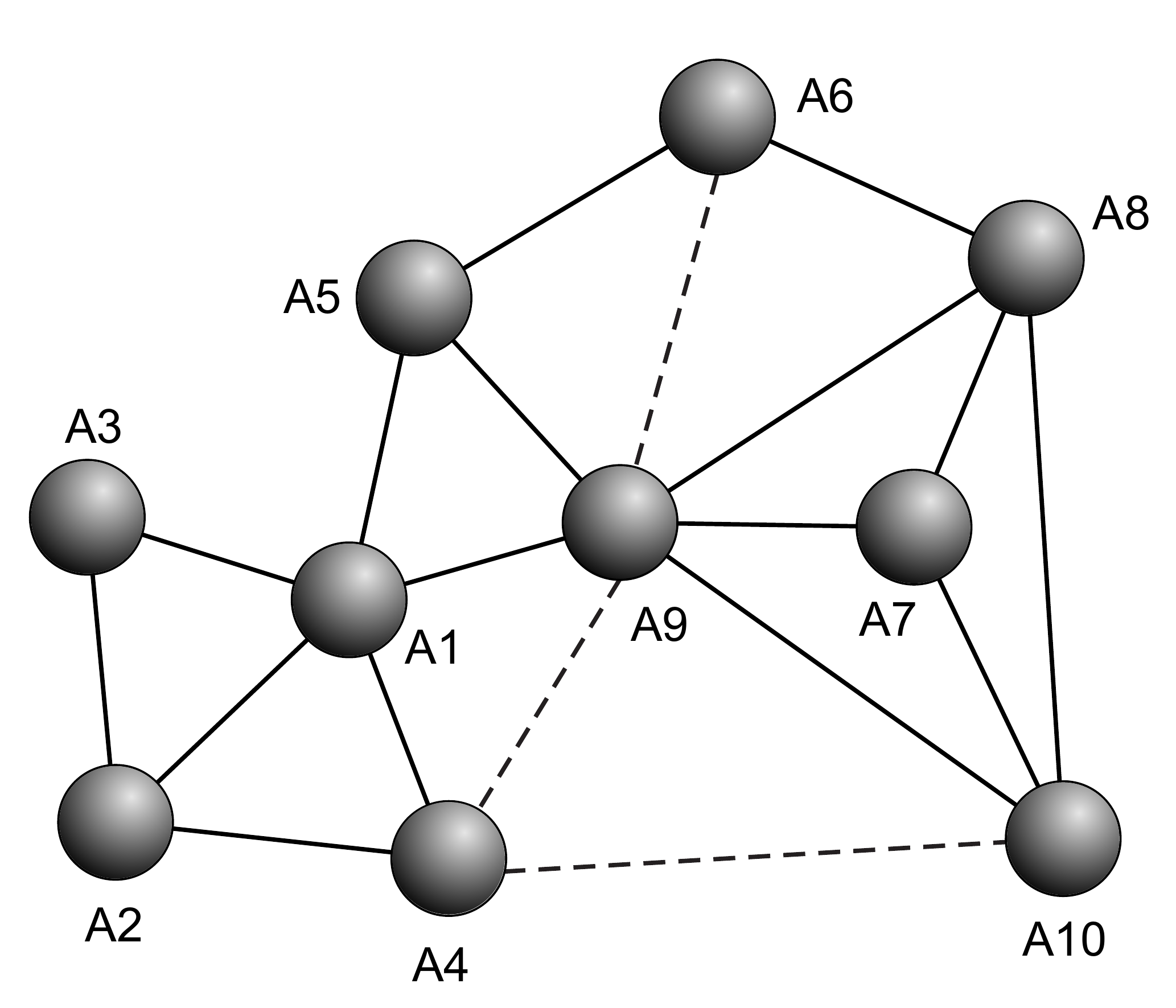}\label{fig:Regular-Lattice-lambda05}}}
    \subfloat[]
    {{\includegraphics[angle=0, width=0.33\textwidth]{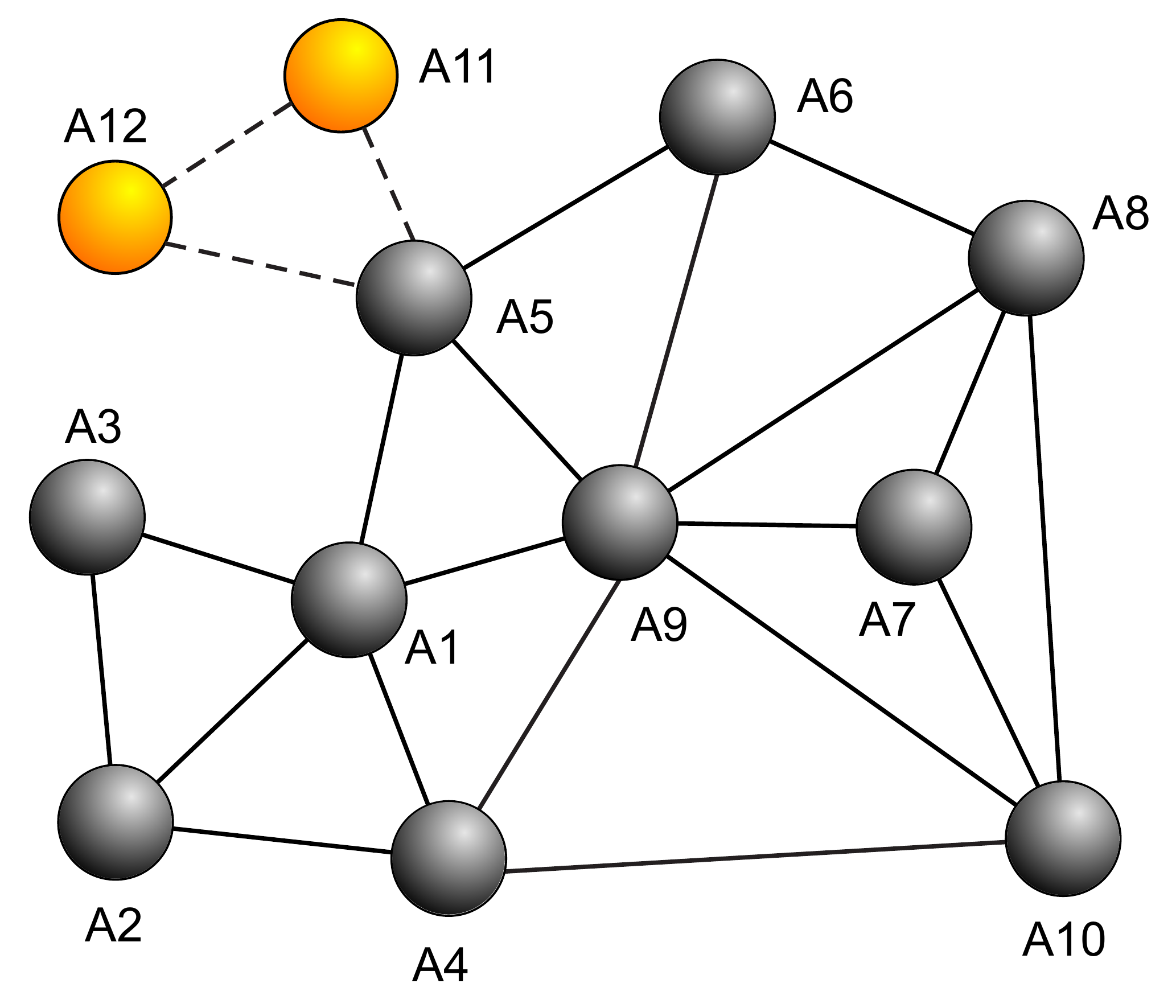}\label{fig:Regular-Lattice-lambda05}}}\\
    \caption{\label{fig.1}
    Example of growing of collaborative networks from $t=t_0$ to $t=t_0 + 2 \Delta t$. The toy database comprises 12 authors and 10 papers.
    (a) Collaborative network at $t = t_0$ built from the following list of 7 papers and 9 authors (A$X$ represents Author $X$): (i) paper 1 (A1, A2 and A3); (ii) paper 2 (A1 and A2); (iii) paper 3 (A1, A2 and A4); (iv) paper 4 (A5 and A6); (v) paper 5 (A7, A8, A9 and A10); (vi) paper 6 (A1, A5 and A9); (vii) paper 7 (A6 and A8). (b) Collaborative network at $t=t_0 + \Delta t$ when paper 8 (A6 and A9) and paper 9 (A9, A4 and A10) are included. New edges are represented as dotted lines. (c) Collaborative network at $t = t_0 + 2 \Delta t$ when paper 10 (A5, A11 and A12) is included. New nodes are represented as orange nodes. }
\end{figure}

The following procedure was applied in order to represent the relationship between authors in a specific topic. Let $A = \{a_{ij}\}$ be the matrix representing the undirected and unweighted network. If authors $i$ and $j$ collaborate on at least one paper from the database, then a link between them is established so that $a_{ij} = 1$. Otherwise, $a_{ij} = 0$. Figure \ref{fig.1} serves as a gist of how the collaborative networks are constructed. Note that at every instant of time, new edges and new nodes might be included in the network.

We built three collaboration networks using the arXiv repository. Each network was built based on papers about an specific topic. {We adopted the criteria employed in ~\citep{model,goodpractices,onthe}: given a keyword, we selected all papers in arXiv which contain this keyword in title or abstract. The keywords chosen were \emph{complex networks}, \emph{graphene} and \emph{topological insulator}. For simplicity's sake we call the respective networks of COMPNET, GRAPHENE, and TOPINSU. These three topics have been chosen for they represent modern topics of current interest in the area of Physics}. Specifically, one network was obtained for each year of the aforementioned networks and the evolution of collaborative groups of authors was studied in terms of the time-varying collaboration networks. Details regarding the networks are given in Table \ref{t:stat}.

\begin{table}
\centering
\caption{\label{t:stat} Database and network statistics. $P$ represents the set of papers, $N$ represents the number of authors and $m$ is the number of edges. $\lambda$ corresponds to the value of the parameter used to fit the exponential growth of the average size of affine groups.}
\begin{tabular}{|c|c|c|c|c|}
 \hline
 Network & $P$ & $N$ & $m$ & $\lambda$\\
 \hline
 COMPNET  & 1316 & 2013 & 5342 & 0.56\\
 GRAPHENE & 4468 & 6490 & 24956 & 1.09\\
 TOPINSU  &  778 & 1436 & 5537 & 1.87\\
 \hline
\end{tabular}
\end{table}

\begin{figure}[!h]
\begin{center}
    \includegraphics[width=1\linewidth]{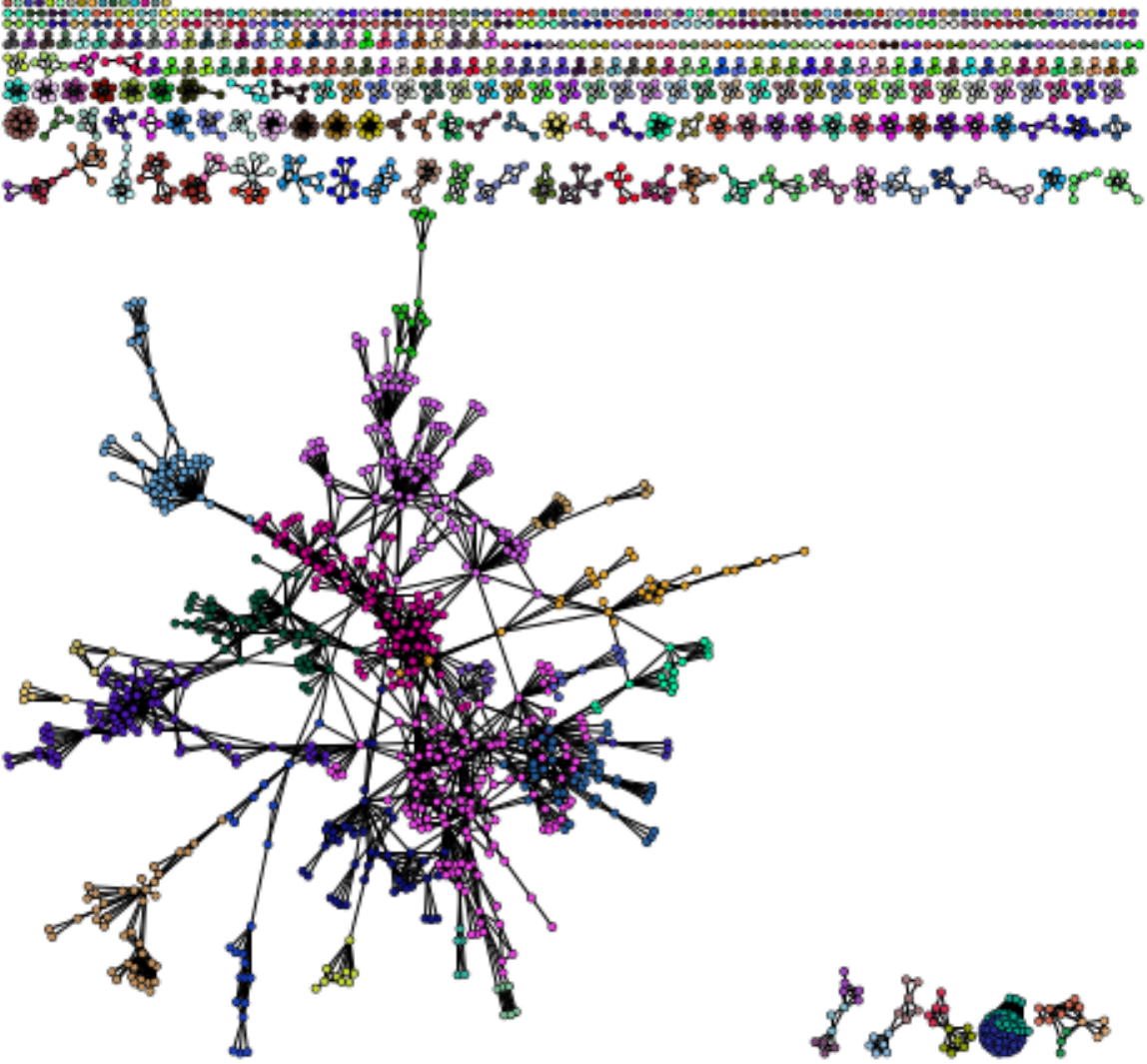}
\end{center}
    \caption{Collaboration network COMPNET of 2011. Communities revealed by the algorithm are represented by different colors of nodes belonging to the giant component. Small components were disregarded. }
    \label{f:2011}
\end{figure}

\section{The affine group}

{Here we define the main concept in this paper, i.e. the \emph{affine group}. For each author $i$ belonging to the set $\Lambda$ of authors  we aim at identifying the subset of authors which are potentially interested in the same subject of research. The most natural choice of authors to belong to the affine group of $i$ are the current or previous collaborators of $i$, i.e., the set $V_i(t) = \{ j \in \Lambda~|~ a_{ij}(t) > 0 \}$. Obviously, authors possibly interested in the same subject may never have collaborated in the past. To consider this case we used the concept of community~\citep{girvan} in networks. A community is a subnetwork (i.e., a group of nodes) that is more densely connected internally than with the other nodes of the network. Formally, a subnetwork is a community if the number of triangles (set of three connected nodes)
is more than $\kappa \sum_i k_i ( k_i - 1 )$, where $\kappa$ is an arbitrary constant~\citep{Seshadhri} and $k_i = \sum_j a_{ij}$.}
This suggests that authors belonging to the same community on average have more affinity than authors belonging to different communities. This might also be inferred from the observation that different communities have particular properties in networks displaying both community structure and assortativity~\citep{newman3}. In order to include the definition of communities in the definition of the affine group, let $C_i(t)$ represent the set of nodes in the same community as $i$ at the instant $t$. Thus, the affine group of a node $i$ at instant $t$, denoted by $G_i(t)$ can be given by the set $G_i(t) = V_i(t) \cup C_i(t)$. To compute $C_i(t)$ we used the algorithm proposed by Newman~\citep{newman3} {(see Appendix A)} to identify the communities for each collaborative network. This algorithm searches for the network partition which maximizes the modularity function $Q$, given as
\begin{equation}
Q = \frac{1}{2m} \sum_{ij} \Bigg{[} a_{ij} - \frac{k_i k_j}{2 m}
\Bigg{]} \delta (c_i, c_j),
\end{equation}
 where $m$ is the number of edges and $\delta (c_i, c_j) = 1$ if nodes $i$ and $j$ belong to the same community and $\delta (c_i, c_j) = 0$ otherwise. We also assume that all the disjoint components with less than ten nodes correspond to a community. Therefore, the algorithm was applied only to the components with more than ten nodes. Figure \ref{f:2011} shows the partition of COMPNET in 2011.

\begin{figure}[h]
\begin{center}
\includegraphics[width=0.7\linewidth]{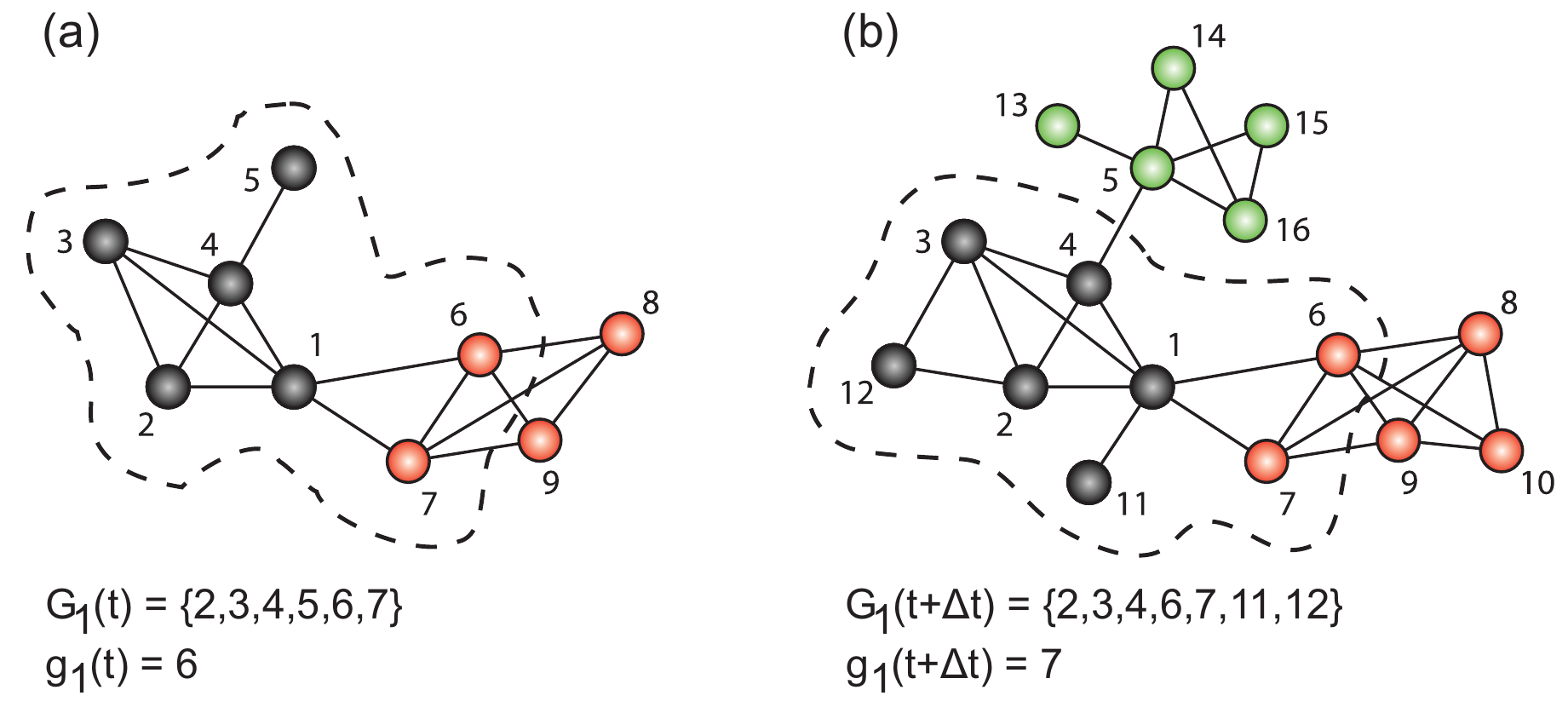}
\end{center}
\caption{Synthetic network where we illustrate the concept of affine group. (a) The network at instant $t$ has 9 nodes and the affine group of node 1, denoted by $G_1(t)$ is given by the set $\{2,3,4,5,6,7\}$, whose size represented by $g_1(t) = 6$. At the instant $t+\Delta t$, the network has 16 nodes and the affine group of node 1 is $G_1(t+\Delta t) = \{2,3,4,6,7,11,12\}$, whose size is $g_1(t+\Delta t)=7$.}
\label{f:synthetic}
\end{figure}

With respect to this definition, the affine group of a researcher is the set of people who could be specially interested in your work, or from who you are mostly interested in. It is important to note that the affine group of a given node changes over time, following the progress of network topology. In order to illustrate this concept, in Figure \ref{f:synthetic} we show a synthetic network at two different time instants as well as the affine group of node $i=1$. In Figure \ref{f:synthetic}(a) one can see that at instant $t$, the network has 9 nodes and the affine group of node $i=1$ is given by $G_1(t)=\{2,3,4,5,6,7\}$. In Figure \ref{f:synthetic}(b), we show the network at instant $t+\Delta t$, which has evolved as consequence of the addition of new authors and new collaborations performed during the time interval $\Delta t$. The affine group of node 1 is now given by $G_1(t+\Delta t)=\{2,3,4,6,7,11,12\}$. Therefore, we can observe that while node 5 no longer belongs to the affine group of node 1, two new nodes (11 and 12) joined this group. The dynamics of how the affine groups evolve along time is the main focus of this paper. It is intrinsically determined by how the authors interact and, consequently, how the communities present in this time-varying collaboration network merge to each other, or split themselves, creating new communities.

\section{Results and Discussion}

\begin{figure}[h]
\begin{center}
   \includegraphics[width=\linewidth]{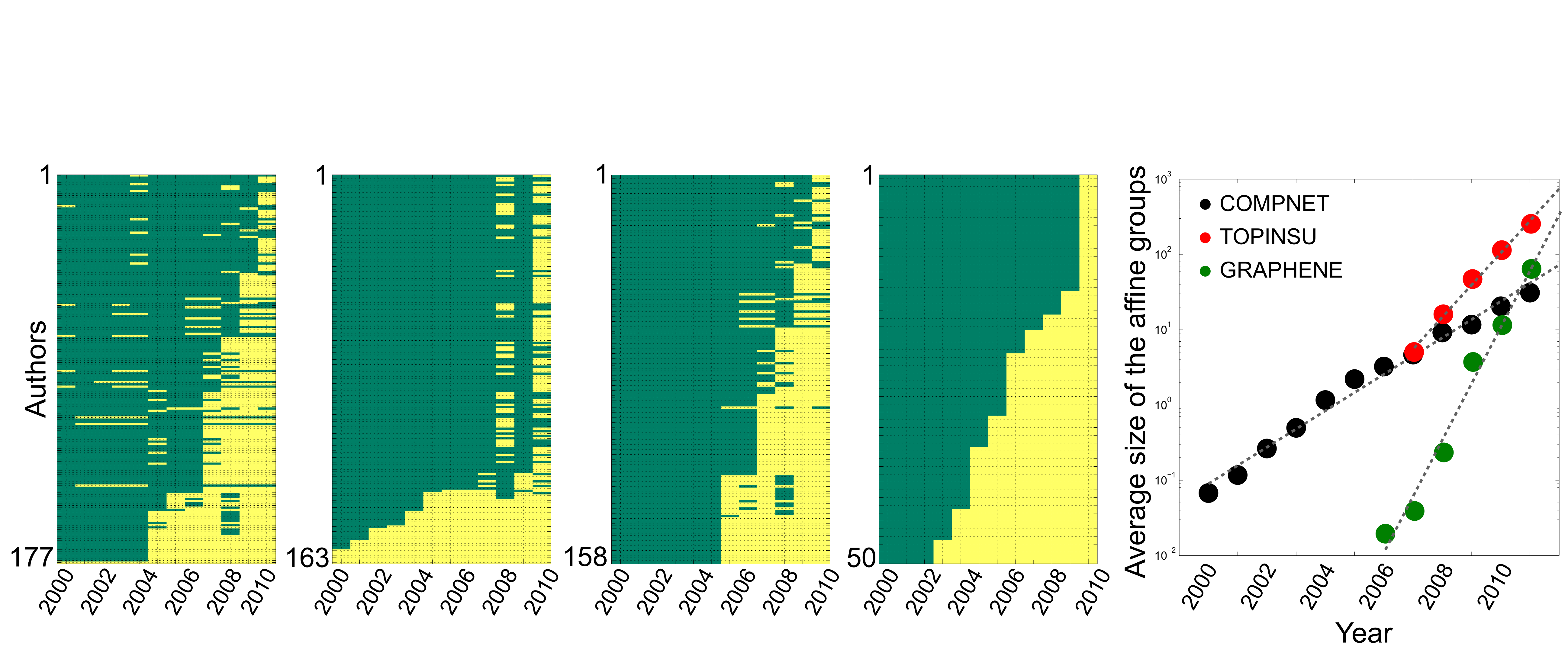}
\end{center}
   \caption{(a-d) Evolution of the affine group of four different authors of the network COMPNET. (e) Evolution of the average size of the affine group for the three collaboration networks considered in this paper. It is important to observe that we also tried to fit the data present in this figure by considering other functions, such as square and cubic polynomial, power-law, and exponential. The exponential fitting presented the minimum Xi-square error.}
\label{f:group}
\end{figure}

In Figure \ref{f:group}(a-d) we show the affine group of four different researchers of the network COMPNET as a function of time. The variable $G_i(t)$ corresponds to each column for different values of $t$. Yellow marks indicate that a given author belongs to the affine group of the reference node at that instant. The rows are sorted from the less active collaboration (top row) to the most active collaboration (bottom row). We see that, in general, the emerging patterns tend to be cumulative, in the sense that since a node $j$ was in the affine group of node $i$ in past, $j$ keeps inside this group in future. As a consequence, we expected that the size of the groups of collaboration tends to increase along time. It is easy to see that the size of a affine group as a function of time, denoted by $g_i(t)$, is given by the sum of the column in Figure \ref{f:group}.

We also studied the average values of $g_i(t)$ averaged over all authors for networks, COMPNET, TOPINSU and GRAPHENE, as a function of time. As the time progresses, the groups increased with average behavior given by an exponential growth. Indeed, when plotting the same curve in monolog scale, as shown in Figure \ref{f:group}(e), we note that the average values are well fitted by the function $\langle g(t)\rangle = g_0e^{\lambda t}$. Figure \ref{f:group}(e) also shows that the exponential behavior is not particular to COMPNET, since both GRAPHENE and TOPINSU also exhibit an exponential growth. In particular, the analysis of $\lambda$ reveals that the average growth of affine groups in the network is faster in the TOPINSU
network. On the other hand, in the COMPNET community, the development of affine groups appears to be more limited and restricted than
the other reasearch fields in Physics. The values of $g_0$ and $\lambda$ which best fit the data for each network are shown in Table \ref{t:stat}.

\begin{figure}[!h]
\begin{center}
\includegraphics[width=0.5\linewidth]{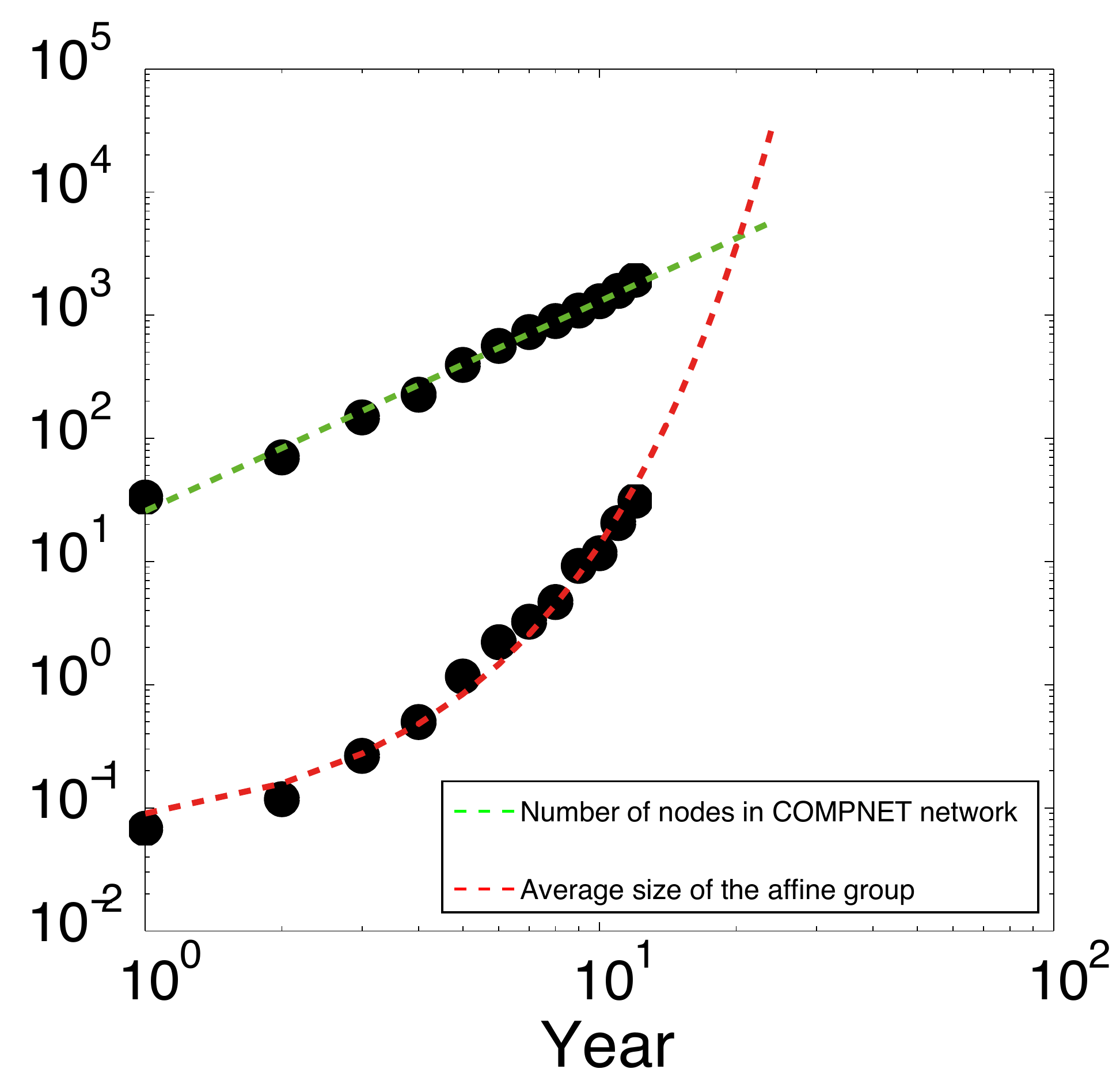}
\end{center}
\caption{Exponential fitting for the average size of the affine group and power-law fitting for the number of authors in the COMPNET ($N(t)\sim t^{1.7}$). The two curves intersect when $t \simeq 20$.}
\label{f:groupevo}
\end{figure}

While the average size of the affine groups grows exponentially, we find out that the number of authors for the three studied networks grows only as power-law along time. This means that there is a tendency of emergence of a single, giant group. In other words, the connectivity patterns of the collaborative networks reveal the imminent emergence of a group of global collaboration. Interestingly, one can predict the emergence of such a group. In the case of COMPNET, there will be a unique affine group around the year 2020, as revealed in Figure \ref{f:groupevo}.

\subsection{Probability for a new connection}

We also evaluated the probability $P$ of a new connection at instant $t+\Delta t$ link two nodes $i$ and $j$, whichever of them was in the affine group of the other. In order to test the robustness of our findings, we report in Table \ref{t:prob} the values of $P$ along time normalized by z-score.
{To compute the z-score, we counted the number $e_r$ of new random edges
that were established inside the affine group. This procedure was repeated 100 times and the average $\langle e_r \rangle$ and standard deviation $\Delta e_r$ were computed. If $\epsilon$ represents the real number of new edges established inside the same affine group, then the z-score is defined as:
\begin{equation} \label{zdef}
    Z = \frac{\epsilon - \langle e_r \rangle}{ \Delta e_r}.
\end{equation}
Note that the normalized version of $P$ given in equation \ref{zdef} quantifies if the number of new links inside affine groups is greater than what would be expected just by chance.}
%The z-score was given by the same number of additional edges, but placed at random over the network. Therefore, .
{The values of Z are given in Table \ref{t:prob}}. With the exception of the network COMPNET in 2004, all observed values are positive. Remarkably, the z-score for TOPINSU and GRAPHENE are particularly high. This means that new collaborations are preferably established within collaborative groups. As such, this pattern could be used, for example, to suggest future collaborations as authors belonging to the same affine group probably shared the same research interests.

\begin{table}
\centering
\caption{{z-score of the probability that a new collaboration to be established within affine groups}. The values are normalized by the probability expected if new connections were established just by chance. Note that almost all values turned out be positive, which confirms that new links are established preferentially within affine groups. \label{t:prob}}
\begin{tabular}{|c|c|c|c|c|c|c|c|c|c|c|c|c|}
    \hline
    Network & 2000 & 2001 & 2002 & 2003 & 2004 & 2005 & 2006 & 2007 & 2008 & 2009 & 2010 & 2011\\
    \hline
    COMPNET  & - & -  & - & - & -0.31 & 4.27 & 10.43 & 4.01 & 9.80 & 2.84 & 6.30 & 0.57\\
    \hline
    TOPINSU  & - & -  & - & - & - & - & - & - & 83.4 & 31.3 & 34.9 & 25.6\\
    \hline
    GRAPHENE  & - & -  & - & - & - & - & - & - & - & - & 20.1 & 29.0\\
    \hline
\end{tabular}
\end{table}

\subsection{Authors classification}

\begin{figure}[!h]
\begin{center}
\includegraphics[width=0.8\linewidth]{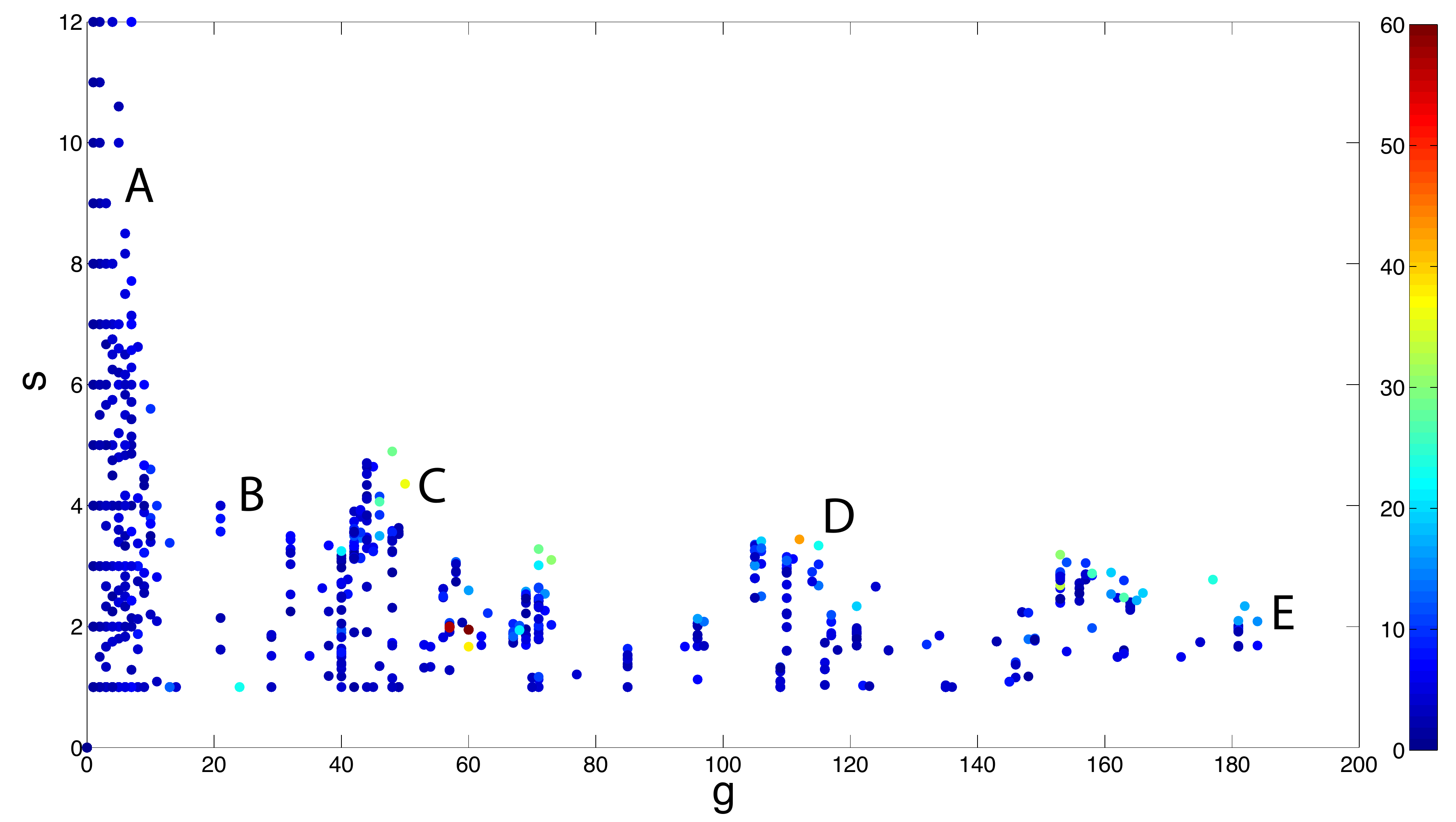}
\end{center}
\caption{Scatter plot between variables $g_{max}$ - \emph{size of affine group} and $s$ - \emph{average duration of collaboration}. Nodes are colored according to the author degree in the network shown in Figure \ref{f:2011}.}
\label{f:plot1}
\end{figure}

As it was shown in Figure \ref{f:group}, different authors tend to have different patterns of collaboration. For instance, the author shown in \ref{f:group}(a) has a kind of intermittent collaborations, which is the opposite of that observed for the author in \ref{f:group}(d). In order to investigate what these patterns tell us about the scientific community, we extracted the following features from the affine group of each author: \emph{i)} $g_{max}$ - the maximum size of the affine group along the years, and \emph{ii)} $s$ - the average size of the period in which authors were in the same affine group. Figure \ref{f:plot1} shows how the authors are distributed according to these two attributes.  It is clear from the overall distribution of authors in this figure that authors that tend to participate in larger affine groups also have shorter collaboration periods.  On the other hand, authors in the region A of Figure \ref{f:group} have small groups, but these groups last for long periods of time. Given that the nodes are colored according to their degree, it also follows from this figure that the previous result is not affected by the number of co-authors.

\section{Conclusions}

The problem of scientific collaboration has been addressed by a large number of works in terms of complex networks.  However, the effect of time has not been usually taken into account.  Actually, the study of time varying models of networks has been restricted to a few works~\citep{perae}. In the present article, we obtained sequences of networks from arXiv parameterized in terms of time, which allowed us to investigate the evolution of collaboration patterns. This was accomplished by defining the affine group of each author and then taking respective measurements regarding the number and duration of pairwise collaborations.  Several interesting results have been obtained.  First, we have that the size of the affine groups follows an exponential law, while the total number of authors grows as power law.  This implies that a single affine group will eventually emerge.  By using extrapolation, we were capable of predicting the date of such an event. Another interesting finding was that researchers tend to exhibit different patterns of collaborations as far as the intermittency is concerned. We mapped these patterns into a 2D space by using the size of the affine groups and the average duration of the collaborations.  It followed from this result that authors that belong to large affine groups tend to have shorter collaboration periods. Another interesting finding is the fact that authors tend to collaborate mainly with colleagues belonging to their affine group, probably because they share the similar research interests. Further works could take into account the intensity of the collaborations (weighted complex networks) as well as investigate the dynamics of migration of authors between groups. Another possibility is to introduce a decay factor for edges weights so that old collaborations are straightforwardly disregarded from the analysis of collaboration patterns.

\section*{Acknowledgements}  	
This work was supported by FAPESP (2010/00927-9, 2010/16310-0 and 2011/50761-2).

\newpage

\appendix
\section{Community detection in complex networks}

{
A feature shared by many real systems modeled as complex networks is the presence of community structure~\citep{girvan}. Nodes of clustered networks tend to organize into groups with many internal connections (above than what would be expected just by chance) and a few external links to the other communities~\citep{newman3}. Social and information networks (see e.g.~\citep{hsiang,chen,ding}) are some examples of networks displaying this type of organization. A large number of recent results suggests that real complex networks may display local properties that are very distinct from the global properties of the entire network, so that the focus on the network as a whole without considering the community structure may overlook many interesting features of the modeled system~\citep{newman3}.
}

{
In the current paper, the method employed to detect communities is based on spectral decomposition~\citep{spectra}. For simplicity's sake, let us consider the case where the network comprises two communities. Let $R$ be the number of edges inter communities, given by:
\begin{equation} \label{req}
	 R = \frac{1}{2} \sum_{G_i \neq G_j} A_{ij},
\end{equation}
where $G_i \neq G_j$  indicates that the sum is considered only if nodes $i$ and $j$ are placed in different communities. The membership of each node is stored in vector $\overrightarrow{s}$  comprising $n$ elements. If node $i$ belong to community $G_1$, then $s_i$ = 1. Otherwise, $s_i = -1$ and $i$ belongs to $G_2$. Note that, if we introduce $\overrightarrow{s}$ in Equation \ref{req}, it can be rewritten as
\begin{equation} \label{r2}
	R = \frac{1}{4} \sum_{ij}(1 - s_i s_j ) A_{ij}.
\end{equation}
Computing the local degree $k_i$ of  $i$ as
\begin{equation}
	 k_i = \sum_{j} A_{ij},
\end{equation}
the total degree of the network is given by
\begin{equation} \label{relacao}
	k = \sum_{ij} A_{ij} = \sum_{i} k_i = \sum_{i} s_i ^ 2 k_i = \sum_{i} s_i s_j k_i \delta_{ij}.
\end{equation}
Replacing equation \ref{relacao} in equation \ref{r2}, one obtains
\begin{equation} \label{r3}
	R = \frac{1}{4} \sum_{ij}(s_i s_j) ( k_i \delta_{ij} -  A_{ij}) = \frac{1}{4} \overrightarrow{s}^{T} L \overrightarrow{s}.
\end{equation}
}
{
Because $R$ represents the number of edges inter-communities, the objective is to minimize this quantity. To do so, we first expand the Laplacian matrix $L$ as a linear combination of the eigenvectors $\overrightarrow{v_i}$:
\begin{equation}
	R = \sum_{i} a_i \overrightarrow{v}_i^{T} L \sum_{j} a_j \overrightarrow{v}_j = \sum_{ij} a_i a_j \lambda_j \delta_{ij} = \sum_{i} a_i^2 \lambda_{i},
\end{equation}
where $\lambda = \lambda_1 \leq \lambda_2 \ldots \leq \lambda_n$ are the eigenvalues associated with $\overrightarrow{v_i}$. In order to minimize $R$, one needs to associate
the highest coefficients $a_i$ to the lowest eigenvalues $\lambda_i$. Thus, the objective is to position $\overrightarrow{s}$ parallel to $\overrightarrow{v_1}$.
Unfortunately, if one followed this optimization strategy a trivial solution would be obtained, since all nodes would belong to a single, giant community. For this reason,
rather than $\overrightarrow{v_1}$, the eigenvector $\overrightarrow{v_2}$ is chosen to set the direction onto $\overrightarrow{s}$ is projected. To minimize the amount of inter communities edges, the best heuristic is to set $s_i = 1$ whenever the $i$-th element of $\overrightarrow{v_2}$ is negative. Analogously, $s_i$ should be set to $s_i = -1$ whenever the  $i$-th element of $\overrightarrow{v_2}$ is positive, minimizing the inner product between $\overrightarrow{s}$ and  $\overrightarrow{v_2}$.
}

\begin{figure*}
\begin{center}
\includegraphics[width=0.55\textwidth]{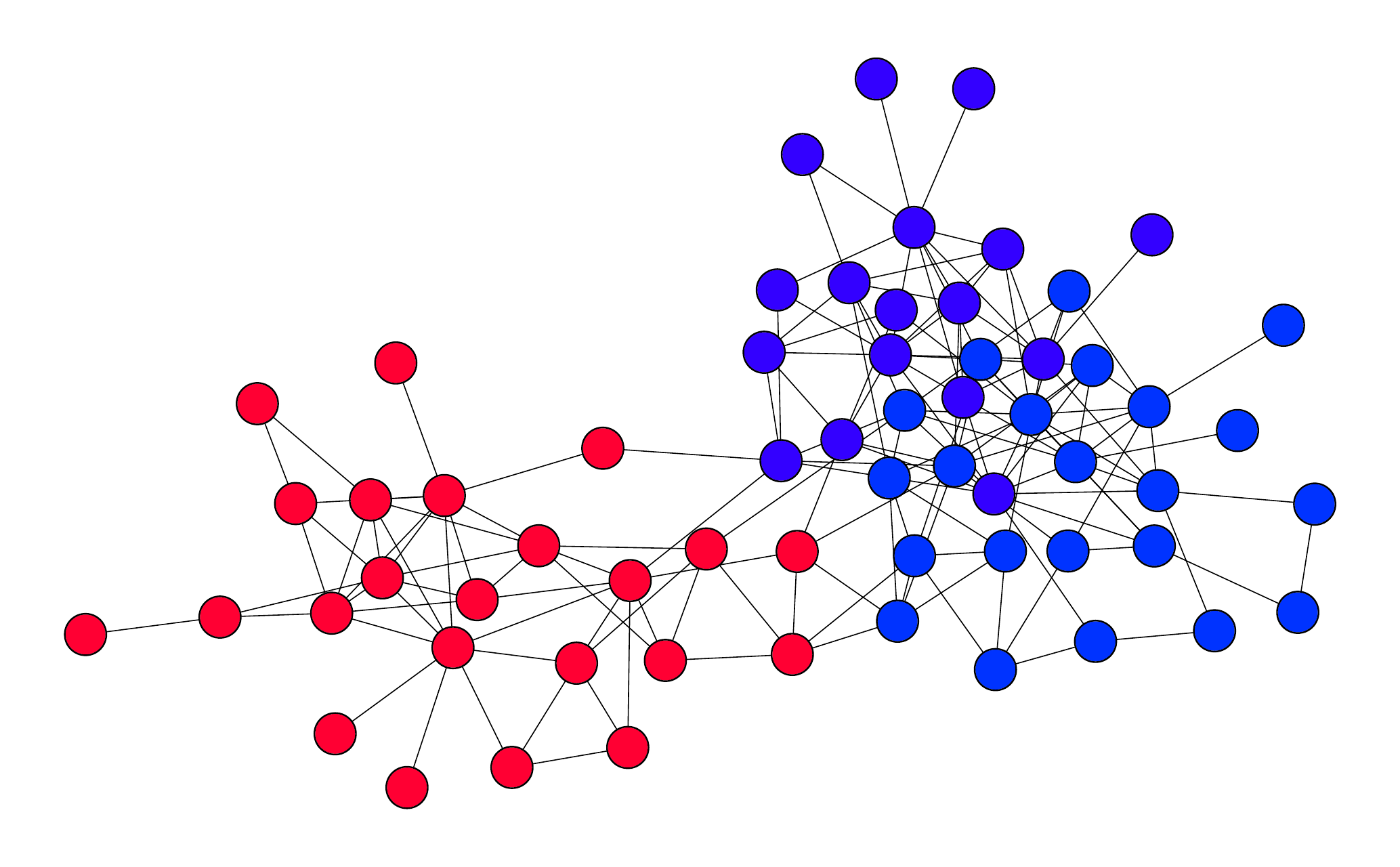}
\end{center}
\caption{Example of partition obtained with the method based on the spectral decomposition of the dolphin social network~\citep{dolphinref}. As expected, each community comprises many intra community edges and a few inter community edges.}
\label{dolphin}
\end{figure*}

{
To exemplify how the algorithm works, we detected two communities in a social network depicted in Figure \ref{dolphin}. As the figure reveals, two partitions were consistently identified. The method employed by \citep{newman3} is very similar to the one described here. Rather than minimizing the number of edges inter communities, \citep{newman3} maximizes the number of edges intra-communities that are above the expected just by chance using the modularity function. As well, they provide methods to identify the potential presence of more than two communities.
}

\end{document}